\documentstyle[prb,aps,twocolumn,amssymb,amsfonts,epsfig]{revtex}

\begin{document}


\title{Spin-phonon interaction and band effects in the high-T$_C$ superconductor HgBa$_2$CuO$_4$.
}

\author
{T. Jarlborg}

\address{D\'epartement de Physique de la Mati\`ere Condens\'ee,
Universit\'e de Gen\`eve, 24 Quai Ernest Ansermet, CH-1211 Gen\`eve 4,
Switzerland} 

\date{\today}
\maketitle

\begin{abstract}
Band calculations show that a stripe-like anti-ferromagnetic spin wave is enforced by
a 'half-breathing' phonon distortion within the
CuO plane of HgBa$_2$CuO$_4$.
This spin-phonon coupling is increased further
by shear distortion and by increased distance between Cu and apical oxygens. 
The effects from spin-phonon coupling are consistent with many observations
in high-T$_C$ materials. Spin-phonon coupling can be important for the mechanism of spin fluctuations and
superconductivity, although the effects are quantitatively weak when using the local density potential.

\end{abstract}


The structures of high-T$_C$ superconductors are essentially 2-dimensional (2-D), containing
flat or buckled CuO-planes. Other 
elements in the structure serve as dopants and mediate the 3-D interaction
between the planes. In addition to the high T$_C$ there are several other
unusual properties characterizing these materials. 
They include the d-wave symmetry of the superconducting gap, existence of a pseudogap, 
isotope effects, anti-ferromagnetic (AFM) fluctuations,
stripe structures and phonon softening, and they all show important dependencies of the
doping \cite{yeh,tall,tsuei,hof,temp,will,mook,zheng,pint}. 
Not only the origin of the high T$_C$ superconductivity
is unknown, but also several of the other unusual
properties in the normal state of these materials are not well understood. 
Many theories for the properties of these superconductors
are discussed in terms of strong correlation \cite{oren}. 
On the other hand, density functional (DF) band calculations can describe the 2-D Fermi surface 
(FS) of the metallic superconducting materials \cite{san},
and it has been
shown that stripe-like spin-wave modulations or certain phonon modes in the CuO planes lead to
gap structures in the one-particle density-of-states (DOS) \cite{tjrap}.

In the present work we study the consequences of interaction between phonons and AFM spin waves
through band calculations for one of the high-T$_C$ materials,
HgBa$_2$CuO$_4$ (HBCO).
This structure has one CuO plane
per formula unit (f.u.) and one FS of cylindrical shape.
The electronic structure calculations are done for ''frozen''
phonon and spin waves by use of the LMTO method and the local spin
density approximation \cite{lda} (LSDA) for supercells of HBCO, including staggered magnetic fields to generate
spin-waves or atomic displacements to model phonons.
The details of the method can be found in refs. \cite{san,tjrap}. 
Previous calculations were made for spin- or phonon
modes of different wave lengths and with different atomic displacements \cite{tjrap}. Here
we consider spin waves together with the 'half-breathing' \cite{shen} oxygen modes along [1,0,0]
in the CuO plane of a cell with 8 f.u. of HBCO,
see fig. 1.  The wavelength of the
spin wave is twice that of the phonon, so that two phonon wavelengths fit into this supercell.
 As is discussed in ref. \cite{tjrap}, gap (or pseudogap) structures appear in the DOS at a band filling of
502 electrons, if this unit cell contains
 one wave length
of a phonon or a spin wave. The gaps appear at 500 electrons, if two wave lengths are contained in the cell. 
This should be compared with the position of the Fermi energy (E$_F$) at a band filling of 504 electrons for 
the undoped cell. In order to account for the doping of 4 holes (overdoped case with 
0.5 holes per f.u.) we use the
virtual crystal approximation (VCA) to adjust the electron filling.  
The distribution of holes follows approximately
the number of holes induced by a rigid  band shift of E$_F$, giving Cu 0.26, 
apical oxygen 0.04 and planar oxygen 
0.08 holes per atom. The nuclear charges are reduced accordingly.

A complete gap appears in the undoped compound (two f.u. of HBCO)
when the staggered magnetic field applied on the Cu-sites
is about 23 mRy. Calculations of the spin wave configurations are made with four
amplitudes of the fields (from 10 to 50 mRy), for 
four different structural modifications within the supercell.
The first case (a) is only with the spin wave, without the phonon. 
In the second case (b) two wave lengths of the half-breathing mode is included along the cell together
with the spin wave, and the phase adjusted so that the nodes of the spin wave coincides with the 
Cu atoms which are 'compressed' between the oxygens, as in fig. 1. 
The displacements of the O-atoms are 0.02$a_0$ ($a_0$ is the lattice constant 7.322 a.u.),
which are realistic for a temperature of 100-200 K \cite{san}.
A third case (c) is like (b), but in a unit cell where the lattice parameter in the ab-plane is reduced
by 0.33 percent, while in the c-direction it is increased by 0.67 percent to conserve the volume.
All atomic positions are scaled by the same c/a-ratio as for the cell dimensions.
This case is motivated by the observation that T$_C$ is lower in thin layers of a high-T$_C$ oxide
grown expitaxially on SrTiO$_3$ (STO) than in the bulk \cite{koll}. As STO has a larger lattice constant than the
high-T$_C$-oxide, it leads to a slight dilatation of the first grown layers. The opposite
distortion in case (c) should make T$_C$ larger, as found experimentally for compressing epitaxial strain \cite{locq}. 
Such results are in 
line with the conclusions from pressure measurements, although oxygen ordering may mask the effect on the
intrinsic T$_C$ \cite{tris}. Mesurements on oxygen rich HBCO-samples find in general positive $dT_C/dP$ \cite{chu}.
The trend from P-measurements on several cuprates seems to be that
T$_C$ is proportional to $a^{-4.5}$ or approximately to the inverse square of the area of
the CuO plane \cite{schill}.  A last case (d) is like (b) except for the fact that the z-distance between Cu
and apical oxygens is increased by 0.01$a_0$. This case is motivated by the reported
correlation between large Cu to apical-O distances (and their overlap integrals)
and large T$_C$'s among a large number
of high-T$_C$ cuprates \cite{pavar}. 

The local Stoner enhancements (S) on the Cu sites are defined as the ratio between exchange splitting
of the Cu-d band and the field energy, $\mu$H.
Without lattice distortion, case (a), the S-enhancement is constant, about 1.45, independent of the field (H).
With the half-breathing phonon in cases (b)-(d) the S-values can be 
extrapolated to about 2.2 for zero field, but the spin wave becomes stiffer at larger H when S are about 1.65. 
In can be noted that the enhancements of case (c) are slightly larger than in case
(b), and that case (d) is more enhanced than (b) for the lowest field. These results indicate already that the phonon
favors a spin wave, and that the shear and increased distance between the Cu and the apical O both increase
the spin-phonon coupling.

One calculation was made in order to test the stability of the spin waves with respect to the 
phase of the phonon.
If the phase of the spin wave in (b) is changed by $\pi$, so the large spin amplitudes are
on the 'compressed' Cu sites, there is clear decrease of the  
spin-phonon coupling. The local $S$-enhancement is 1.35, much lower than the results from the
other cases, even lower than the case (a) without phonon.

In view of the results for separate spin waves and phonons, it is somewhat surprising to find that the
gap for the co-existing spin- and phonon waves of fig. 1 appears at the band filling of 500 electrons.
This means that the gap coming from the charge
modulation of the phonon is enforced by the spin wave. Furthermore, the spin wave is
stronger when it coexists with the phonon.  The phonon alone gives a faint hint of a pseudogap at the band
filling of 500 electrons. But when the spin wave is added, the gap is stronger, and is almost complete
for the largest fields.  
This can be understood from the
fact that the charge of an unpolarized Cu atom is not the same as the charge of a polarized one.
The latter, the sum of majority and minority spin distribution, makes
the non-spin dependent part of the potential different from that on
unpolarized Cu and hence the charge modulation is enforced.
These results point also towards close interactions between phonon and spin waves.

The coupling constant of spin fluctuations is calculated 
by the same method as for Fe, Co and Pd \cite{tjfe}, which is based on an analogy between $\Delta r$,
the sum of the amplitudes of atomic displacements in the case 
of electron-phonon coupling, and $m$, the sum of the amplitude of the magnetic moments 
for spin waves. The matrix element for the change in potential $ < \delta V > $ is calculated 
from the FS average of the difference of band energies $\epsilon^m_k - \epsilon^0_k$ between the cases
with and without magnetization. The result is closely linear in $m$, $ < \delta V > = I \cdot m$.
The difference in total energy between the configuration with moment $m$ and at zero moment,
$E_m - E_0$ depends almost quadratically on $m$, $E_m - E_0 = F m^2$, similar to the dependence
of the total energy with $\Delta r$ for harmonic phonons. By taking the derivatives with respect
to $m$ we have in the harmonic case \cite{tjfe},
\begin{equation}
\label{eq:lamsf}
\lambda_{sf} = N I^2/2F 
\end{equation}
where $N$ is the DOS at E$_F$ ($\sim$105 states/cell). 
A BCS-like equation is sufficient to discuss variations of T$_C$ at this stage;
$T_C \sim \omega_{sf} exp(-1/\lambda_{sf})$
which does not yet contain the separate effects from s-, p-wave,
or electron-phonon coupling \cite{fay}. Neither the amplitude of the prefactor $\omega_{sf}$
is known for the spin-phonon coupled system, but it can be concluded that a large T$_C$
requires a large $\lambda_{sf}$. 

All calculations are made using 24 irreducible k-points, and the matrix elements $I^2$ 
are averaged over the eigenvalues near $E_F$ for these points. 
The total energy results converge more slowly than the S-factors, moments or matrix elements,
and many self-consistent iterations are required.
Table I shows the results of $I^2$, $F$ and $\lambda_{sf}$ evaluated at the largest field
for the 4 cases, (a)-(d). The values for $F$ evaluated at the lower fields are almost identical for the
3 cases containing the phonon, (b)-(d), i.e. the harmonicity is almost perfect. For case (a) however,
$F$ is strikingly non-harmonic. $F$ evaluated at the lowest field is about 5 times larger
than the value given in Table I,
indicating that case (a)
is not ready for spin waves. Only at large applied field the system is forced to become magnetically softer.
Thus, the high field value in Table I represents an upper limit of $\lambda_{sf}$ for case (a),
not taken in the limit $m \rightarrow 0$. 
 The linear $m$-dependence of $I$ is respected faily well in view of the limited statistics coming
from the k-point mesh. Still there is a tendency that the values are 10-30 percent larger at the 
highest field compared to at the lowest one.
Both case (c) and (d) give stronger coupling than case (b) confirming the expectation that smaller
CuO areas and/or large distance between
apical oxygen and the CuO plane are indeed important for a T$_C$-mechanism based on spin fluctuations.
The relative differences between cases (a)-(d), showing that when a gap is opened easily (and the matrix elements
are large) in the latter cases (b)-(d), the total energy increases more slowly with field. Such
cases are magnetically 'soft', and the coupling calculated from eq. \ref{eq:lamsf}, is largest.
It can be noted that $\lambda_{sf}$ varies considerably among the different k-points,  
typically up to a factor 3 compared to the FS average in all cases (a)-(d)
 \cite{tjrap}.

 The value of $\lambda_{sf}$, of the order 0.25 for the most magnetically soft cases,
 is not large enough to explain a high T$_C$.  Therefore, no quantitative discussion
of T$_C$ can be done yet. The problem is most likely connected with the DF-LSDA potential,
which is known to be unable to describe many AFM ordered insulating oxides. As was said, the calculations
for the undoped HBCO need a staggered field of about $\pm 23$ mRy to form a true gap, while in reality 
no field should be needed. It could be argued that the system is magnetically
much softer than what is found in LSDA, so that some part of the applied field
should be intrinsic in order to correct for the error in the LSDA, and a larger $\lambda_{sf}$
can be imagined. Strong correlation, self-interaction corrections
or non-locality have been proposed to improve the DF results. 
 The CuO-plane is very close packed and the charge density has a deep minimum in the
binding region between Cu and O.
Exchange and correlation (xc) is larger than in LDA when
the charge density has a minimum \cite{tjpot}, because the xc-hole is more localized than for
the constant density taken at the minimum of real density. Non-locality can be approached by 
taking the average density within
$r_s$, the electron gas parameter, instead of the local density in the evaluation of the potential.
A calculations based on such a correction of the LDA potential requires lower critical field ($\sim$ 15 mRy) 
to open a gap in the undoped case.  When the same correction is applied to case (b) it leads to 30-50
percent increase of $I^2$. Such results are not ab-initio, but they are useful to demonstrate that magnetic 
softness correlates with larger coupling.
The reason for the larger magnetic enhancement and larger coupling can be found in the position of the O-p
band relative to E$_F$. It turns out that the coupling is larger when the 
top of the O-p band is closer to E$_F$,
as in the model calculation
with the corrected potential or as in the LSDA results for case (c) compared to case (b).
The orbital overlap between atomic neighbors is small at the top of the band, and this can reduce
the hybridization energy and make AFM configurations more stable. 
Improved non-locality beyond LSDA or gradient corrected potentials will
be important for quantitative results of nearly AFM systems.
In the absence of
a well tested theory for improved DF calculations we can, in the following, only make a qualitative
discussion of the consequences of spin-phonon coupling.

It is known that non-magnetic substitutions like Zn (on Cu sites) suppress T$_C$ more rapidly 
than magnetic ones like Ni in hole doped copper oxides \cite{yeh}.
Two calculations with spin waves in the undistorted case (a) were made
with one Cu-site replaced by either a Ni or a Zn atom. Table II shows the resulting
moments for the two cases and the case without impurity, when the field amplitude is 20 mRy
in all cases. The d-band in Zn is filled and an applied field gives a spin splitting, but no local moment.
 It is seen that the Zn impurity perturbs the AF order
even on distant Cu, and totally there is a sizable ferromagnetic moment. 
Ni acquires a rather strong moment, but the spin distribution on the remaining Cu
sites is not too far from the case without impurity.
These results indicate that AFM spin waves and possibly superconductivity
mediated by spin fluctuations are more sensitive to Zn than to Ni impurities, as the observed trends 
for $T_C$ \cite{yeh}.

The mechanism of spin-phonon coupling is consistent with the fact that both atomic
vibrations and magnetic fluctuations, observed via isotope effects and 
spin polarization, play a role for the superconducting state.
 So far we only consider the half-breathing mode \cite{shen,mcqueen}. This mode 
 can be modeled by 
a longitudinal wave along a 1D-CuO chain for
a two atomic chain with masses $M_{Cu}$ and $M_O$ \cite{zim}. 
The phonon energies at the zone boundary are $\sqrt(2K/M_O)$ and
$\sqrt(2K/M_{Cu})$, where $K$ is a force constant. 
The upper state concerns movements of the O-atoms as in the frozen
phonon calculations, with very different moments on 'compressed' and 'dilated' Cu.
(The lower state is for a similar movement of 
the Cu atoms. This movement is less crucial for Cu moments, because the distance to the oxygens
are just increased on one side and decreased on the other side of each Cu.) 
The square of the vibrational amplitude, $u^2$=$3\hbar\omega/2K$, with $K=M\omega^2$, 
gives $u^2 \sim M^{-1/2}$, when T is low compared to the Debye temperature \cite{san}. 
Thus, a smaller $M_O$ tends to
increase the vibrational amplitudes of the oxygens, which favors magnetic fluctuations 
and larger $\lambda_{sf}$. The mass dependence might seem small, but there is a feedback from 
the force constant, $K$.
An increased $u^2$ will enforce the spin wave and decrease the total energy, so that the phonon appears
softer, i.e. a lower $K$. The effect is to increase the vibrational amplitude, the spin amplitude and
$\lambda_{sf}$.  
Isotope shifts of T$_C$ are complex, 
since a large T$_C$ requires that both $\omega_{sf}$ 
and $\lambda_{sf}$ are large, while in general a
large $\lambda_{sf}$ leads to a small $\omega_{sf}$ and vice-versa \cite{tjfe}.
Larger isotope shifts 
can be expected for the pseudogap, mainly because the gap depends directly on the 
magnetic and vibrational amplitudes. 
Furthermore it exists at
larger T, where $u^2$ and the effects of feedback via $K$ are probably larger. 
Results from inelastic neutron scattering show larger isotope shifts of the pseudogap
than of $T_C$ \cite{temp}.
As mentioned above,
vibrations of Cu are not so important for the Cu moments and $\lambda_{sf}$ in this mechanism. Increased
amplitudes of Cu-vibrations could even interfere with the O-vibrations to make the effects from
the half-breathing mode less effective, and lead to negative isotope effects, as are
found in some cases for $T_C$ \cite{will}.

Another feature of high-T$_C$ compounds is the softening of zone boundary phonons
in doped materials \cite{pint,mcqueen}. There is a correlation between doping and wave lengths of the modulations
\cite{tjrap}, so that only if the gap opens at the correct $E_F$ of the doped system, there will
be a lowering of the total energy. Thus, the O-phonon 
mode in fig. 1, which corresponds precisely to the zone-boundary phonon in the simple model above,
appears to be softer than if no spin wave existed.
If the system is undoped, $E_F$ will be at a higher energy, above the gap caused by this
spin-phonon mode, and no gain in total energy will soften the phonon.

The properties of electron doped cuprates seem quite different from the hole doped ones, with no clear
indications of pseudogaps or d-wave symmetry of the superconducting gap \cite{yeh}.
The band results show more often that the strongest gaps
appear below the position of E$_F$ for the undoped material \cite{tjrap}  
(A few exceptions are found for separate spin and phonon waves). A possible 
explanation is that the charge difference between polarized and non-polarized Cu sites is smaller  
when $E_F$ is moving away from the Cu-d band.

In conclusion, the band results on supercells of HBCO along the CuO bond direction show that spin waves
interact strongly with "half-breathing" phonons. Static stripes can be viewed as instabilities  
of the phonons due to interaction with the spin wave. Soft spin-phonon waves, before the instability limit,
imply possibilities of superconductivity mediated by spin fluctuations. Many properties of high-T$_C$ cuprates, 
such as isotope shifts, phonon softening in doped systems, magnetic fluctuations in the superconducting state,
existence of pseudogaps at E$_F$, sensitivity to magnetic and non-magetic impurities, and
variations with pressure or strain can be understood 
qualitatively within the mechanism of spin-phonon coupling.  
 
Acknowledgment. I am grateful to E. Koller, S. Reymond and H. Wilhelm for various discussions.



\begin{table}
\caption{Matrix element $I^2$ ($10^{-3}Ry^2/ \mu_B^2$), parameter for the total energy 
difference $F$ (mRy/cell/$\mu_B^2$), and $\lambda_{sf}$
 for the four different cases (a)-(d) described in the text.
}
\begin{tabular}{ccccc}     
  & case (a) & case (b) & case (c) & case (d)    \\
\tableline
$I^2$ & 0.05 & 0.12 & 0.14 & 0.15 \\
$F$ & 77 & 33 & 30 & 30 \\
$\lambda_{sf}$ & 0.04 & 0.19 & 0.24 & 0.27 \\
\end{tabular}
\end{table}
\begin{table}
\caption{Magnetic moment per site (site numbers in fig. 1) for case (a) without and with one Ni or Zn
impurity on site 1 for a field amplitude of 20 mRy. The total moments per cell are 0.00, 0.47 and -1.04,
respectively.
}
\begin{tabular}{cccc}     
 site & all Cu & Ni on site 1 & Zn on site 1   \\
\tableline
1 & 0.36 & 0.79 & 0.01 \\
2 & -0.36 & -0.28 & -0.53 \\
3 &  0.0 & -0.01 & -0.06 \\
4 &  0.0 & -0.01 & -0.05 \\
5 & -0.36 & -0.27 & -0.34 \\
6 &  0.36 &  0.27 & -0.23 \\
7 & 0.00 & -0.01 & -0.06 \\
8 & 0.00 & -0.01 & -0.04 \\
\end{tabular}
\end{table}

\begin{figure}[tb!]

\leavevmode\begin{center}\epsfxsize8.6cm\epsfbox{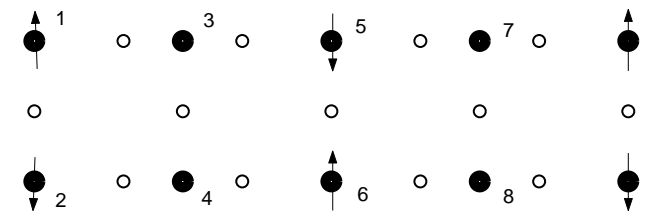}\end{center}
\caption{
Schematic view of a CuO plane containing a 'half-breathing' O-phonon (wave length 2$a_0$) and a spin wave
(wave length 4$a_0$)  along [1,0,0]. 
The filled, numbered circles are Cu-sites and open circles the plane oxygens.
The arrows indicate up- and down-moments on Cu. The displacements of the O-atoms 
are exaggerated for visibility, in the calculations (cases b-d) they are 0.02$a_0$.}
\end{figure}

\end{document}